\documentclass[12pt,letterpaper]{paper}
\usepackage[margin=1in]{geometry}
\usepackage{url}
\usepackage{indentfirst} 

\usepackage{amsmath,amssymb}

\usepackage[font=small,labelfont=bf]{caption}
\usepackage[labelsep=period]{caption}
\usepackage{threeparttable} 
\usepackage{afterpage} 
\usepackage{pdflscape} 
\usepackage{tablefootnote}
\usepackage{lscape}

\usepackage{multirow}
\usepackage{enumitem} 
\usepackage{array}

\usepackage{tikz}
\usetikzlibrary{arrows, positioning}
\usepackage{graphicx}
\newcolumntype{P}[1]{>{\centering\arraybackslash}p{#1}}
\usepackage{dcolumn}
\newcolumntype{d}[1]{D{.}{.}{#1}}
\usepackage{setspace} 
\usepackage{ragged2e} 
\usepackage{fancyhdr}
\usepackage{hyphenat} 
\usepackage{cite} 
\RaggedRight 
\usepackage{pdfpages}

\begin{document}
\begin{minipage}{1\textwidth}
\large 
\centering \textbf{Sensitivity Analysis for Random Measurement Error using Regression Calibration and Simulation-Extrapolation}\\[4ex]
\small
\begin{flushleft}
Linda Nab and Rolf H.H. Groenwold\\[4ex]
Correspondence to Linda Nab MSc, Department of Clinical Epidemiology, Leiden University Medical Center, Postzone C7-P,
P.O. Box 9600, 2300 RC Leiden,
the Netherlands (email: \url{l.nab@lumc.nl})\\[4ex]
Author affliations: Department of Clinical Epidemiology, Leiden University Medical Center, Leiden, the Netherlands (Linda Nab, and Rolf H.H. Groenwold); and Department of Biomedical Data Sciences, Leiden University Medical Center, Leiden, the Netherlands (Rolf H.H. Groenwold). \\[4ex]
This study was supported by grants from the Netherlands Organization for Scientific Research (ZonMW-Vidi project 917.16.430) and Leiden University Medical Center.\\[4ex]
Conflicts of interest: none declared.\\[4ex]
Running head: Sensitivity analysis for random measurement error\\[4ex]
Data and code availability: The data and code used for the simulation study have been made publicly and can be accessed via \texttt{www.github.com/LindaNab/simexvsmecor}.
\end{flushleft}
\end{minipage}

\thispagestyle{empty}

\newpage
\pagestyle{fancy} 
\fancyhf{} 
\renewcommand{\headrulewidth}{0pt}
\fancyfoot[R]{\thepage}

\begin{abstract}
\noindent (200/200)\\
\textbf{Objective.} Sensitivity analysis for measurement error can be applied in the absence of validation data by means of regression calibration and simulation-extrapolation. These have not been compared for this purpose.\\
\textbf{Study Design and Setting.} A simulation study was conducted comparing the performance of regression calibration and simulation-extrapolation in a multivariable model. The performance of the two methods was evaluated in terms of bias, mean squared error (MSE) and confidence interval coverage, for ranging reliability of the error-prone measurement (0.2-0.9), sample size (125-1,000), number of replicates (2-10), and R-squared (0.03-0.75). It was assumed that no validation data were available about the error-free measures, while measurement error variance was correctly estimated.\\
\textbf{Results.} In various scenarios, regression calibration was unbiased while simulation-extrapolation was biased: median bias was $1.4\%$ (interquartile range (IQR): $0.8$;$2\%$), and $-12.8\%$ (IQR: $-13.2$;$-11.0\%$), respectively. A small gain in efficiency was observed for simulation-extrapolation (median MSE: $0.005$, IQR: $0.004$;$0.006$) versus regression calibration (median MSE: $0.006$, IQR: $0.004$;$0.007$). Confidence interval coverage was at the nominal level of 95\% for regression calibration, and smaller than $95\%$ for simulation-extrapolation (median coverage: $92\%$, IQR: $85$;$94\%$).\\
\textbf{Conclusion.} In the absence of validation data, the use of regression calibration is recommended for sensitivity analysis for measurement error. 
\end{abstract}
\newpage
\section{INTRODUCTION}
Measurement error is common in biomedical research but often ignored \cite{Brakenhoff2018MeasurementReview, Shaw2018EpidemiologicRecommendations}. When ignored, measurement error can lead to considerable biases in exposure-outcome associations \cite{Keogh2020STRATOSAdjustment}. Random measurement error in the exposure variable is common in various domains of epidemiology \cite{Frazer2019VariabilityStudy,Perrier2016Within-subjectStudies, Brunekreef1987VariabilityEpidemiology}. Random measurement error in an exposure variable introduces bias in the exposure-outcome association, which is sometimes referred to as attenuation bias \cite{Spearman1904TheThings} or regression dilution bias \cite{Frost2000CorrectingVariable., Hutcheon2010RandomBias}.

Various methods for measurement error correction are available \cite{Armstrong1985MeasurementModel,Bartlett2018BayesianCalibration,Buonaccorsi2010MeasurementApplications, Carroll2006, Cole2006Multiple-imputationCorrection, Cook1994Simulation-ExtrapolationModels, Gustafson2004MeasurementAdjustments., Fuller1987MeasurementModels, Keogh2014AEpidemiology}, yet application of these methods is rare in biomedical research \cite{Keogh2020STRATOSAdjustment}. One possible barrier is the necessity of for instance validation data. Validation data contain extra measurements taken to validate error-prone measurements, which are often unavailable \cite{Perrier2016Within-subjectStudies}. Validation data can be used to estimate the measurement error model and its parameters, and subsequently used for measurement error correction. 

In the absence of validation data, regression calibration \cite{Gleser1991MeasurementModels, Carroll1990ApproximatePredictors} and simulation-extrapolation \cite{Stefanski1995Simulation-extrapolation:Jackknife}, for instance, can be applied to correct for random exposure measurement error. Both methods only require assumptions about the variance of the random measurement error, for example based on literature or expert knowledge. Regression calibration in the absence of validation data is available in the R \cite{Rsoftware} package \texttt{mecor} for measurement error correction \cite{Nab2021Mecor:Outcome}, that implements the regression calibration described by Rosner et al. \cite{Rosner1992CorrectionError}.  Alternatively, simulation-extrapolation is easy to use due to its implementation in the R package \texttt{simex} \cite{Lederer1997AMCSIMEX} and the \texttt{simex} procedure \cite{Hardin2003TheError} in Stata \cite{StataCorp2019Stata16}.

Simulation-extrapolation and regression calibration have been compared in simulation studies for scenarios where replicate measures of the error-prone exposure were available \cite{Perrier2016Within-subjectStudies,Batistatou2012PerformanceData}. The results of the studies by Perrier et al. \cite{Perrier2016Within-subjectStudies} and Batistatou et al. \cite{Batistatou2012PerformanceData} were consistent and showed that, regression calibration and simulation-extrapolation reduced bias compared to when no measurement error correction was applied or when the replicate exposure measures were pooled. It was also shown that application of simulation-extrapolation generally produced more biased effect estimates than regression calibration, especially when the reliability of the error-prone measure was low.

Perrier et al. \cite{Perrier2016Within-subjectStudies} and Batistatou et al. \cite{Batistatou2012PerformanceData} only studied a limited number of scenarios, e.g., large sample sizes, limited range of reliability of the error-prone measure, and univariable models only. Further investigation is needed of the performance of regression calibration and simulation-extrapolation in more complex settings, as typically found in epidemiologic research. Moreover, since the simulation studies by Perrier et al. and Batistatou et al. focused on settings where replicate measures were available, we aim to research how their results translate to settings were no replicate measures, but only an estimate of the measurement error variance is available. The quantification of the performance of the two methods in this broader range of settings is used as the input for a framework guiding the application of measurement error correction in the absence of validation data.

This paper is structured as follows. Section 2 reviews and applies regression calibration and simulation-extrapolation by using two motivating examples of error-prone systolic blood pressure measurements. In section 3, a simulation study is described that aims to compare regression calibration and simulation-extrapolation, and results from the simulation study are shown. Section 4 introduces a framework for conducting sensitivity analysis, also known as quantitative bias analysis \cite{Lash2009ApplyingData}, for random measurement error by means of regression calibration and simulation-extrapolation. We conclude with a discussion of our results and recommendations in section 5.  

\section{MOTIVATING EXAMPLE}
To demonstrate the use of regression calibration and simulation-extrapolation to correct for random measurement error, we used an example about the association between blood pressure (systolic blood pressure) and kidney function (serum creatinine). \bigskip

\noindent \textit{Example 1: Blood pressure and kidney function in adults} \\
For the first example, we used data of the 2015-2016 cycle of the National Health And Nutrition Examination Survey \cite{NHANES}. Given natural variation of blood pressure within individuals, a single measure of blood pressure often does not reflect the true level of blood pressure. Therefore, in epidemiological studies, often multiple blood pressure measurements are obtained. In NHANES, three consecutive systolic BP measurements were taken, after resting quietly in a seated position for 5 minutes. The serum creatinine concentration was determined once. Demographic information was collected using the family and sample person demographics questionnaires in the home, by trained interviewers. For this analysis, pregnant women, individuals receiving dialysis in the past 12 months or indicating that that they had been told they had weak/failing kidneys were excluded from analysis. Additionally, all individuals with a systolic blood pressure greater than 180 mmHg and serum creatinine greater than 200 $\mu$mol/L were excluded.

First, the association between systolic blood pressure and serum creatinine was determined by only using the first systolic blood pressure measure. The association was adjusted for age. It was found that an increase of 10 mmHg in systolic blood pressure was associated with a 0.87 $\mu$mol/L (95\% confidence interval (CI): 0.51-1.22) increase in serum creatinine (Table 1). However, the random measurement error in the first systolic blood pressure measure was not considered here. In the NHANES data, the within individual variation of a systolic blood pressure measure was on average 13.1 mmHg, which was obtained by using the three consecutive systolic blood pressure measures. The within individual variation of 13.1 mmHg was subsequently used to correct for the measurement error in the first systolic blood pressure measure using regression calibration and simulation-extrapolation, while adjusting for age. An explanation of the two methods can be found in subsection \ref{sec:reviewrcsimex}. Using regression calibration, we found that an increase of 10 mmHg in systolic blood pressure was associated with an 0.93 $\mu$mol/L (CI: 0.54-1.32) increase in serum creatinine (Table 1). Using simulation extrapolation, we found that an increase of 10 mmHg in systolic blood pressure was associated with an 0.93 $\mu$mol/L (CI: 0.50-1.36) increase in serum creatinine (Table 1). \medskip

\noindent \textit{Example 2: Blood pressure and kidney function in pregnant women}\\
For the second example, we used data of retrospective records of all women who attended a tertiary maternity hospital pregnancy day assessment clinic over a 6-month period in 2014 in Australia \cite{McCarthy2015EffectivenessHospital}. Care always included serial, manual blood pressure measurements every 30 min by registered midwives using aneroid sphygmomanometers \cite{McCarthy2015EffectivenessHospital}. Serum creatinine and demographic data were obtained using routinely collected data. One woman with a serum creatinine level lower than 10 $\mu$mol/L was excluded from the analysis.  

First, the association between systolic blood pressure and serum creatinine was determined by only using the systolic blood pressure measure obtained after 30 min. The association was adjusted for age. We found that an increase of 10 mmHg in systolic blood pressure was associated with a 1.18 $\mu$mol/L (CI: 0.14 - 2.23) increase in serum creatinine (Table 1). In this analysis, the random measurement error in the single systolic blood pressure measurement was not taken into account. Using the four consecutive blood pressure measurements (obtained after 30, 60, 90 and 120 minutes), it was found that the within individual variation of the systolic blood pressure measures was on average 48.3 mmHg. The within individual variation of 48.3 mmHg was subsequently used to correct for the measurement error in the single systolic blood pressure measurement using regression calibration and simulation-extrapolation, while adjusting for age. Using regression calibration, we found that an increase of 10 mmHg was associated with a 2.04 $\mu$mol/L (CI: 0.23-3.91) increase in serum creatinine (Table 1). Using simulation-extrapolation, we found that an increase of 10 mmHg was associated with a 1.67 $\mu$mol/L (CI: 0.01-3.25) increase in serum creatinine (Table 1).

\begin{table}[ht]
    \centering
    \caption{Effect estimates (95\% confidence intervals) of the association between blood pressure (systolic blood pressure, per 10 mmHg) and kidney function (serum creatinine, $\mu$mol/L). The uncorrected effect estimates are obtained using the first blood pressure measure only, the corrected estimates are obtained by using the two consecutive measurements (example 1) and the three consecutive measurments (example 2).}
    \label{tab:motex}
    \begin{threeparttable}
    \begin{tabular}{|l|l|l|l|l|}
    \hline
         Study & Uncorrected & Regression Calibration & Simulation-extrapolation  \\
         \hline
         Example 1 & 0.87 (0.51-1.22) & 0.93 (0.54-1.32) & 0.93 (0.50-1.36)\\
         \hline
         Example 2 & 1.18 (0.14-2.23) & 2.04 (0.23-3.91) & 1.67 (0.01 - 3.25)\\
         \hline
    \end{tabular}
      \begin{tablenotes}[flushleft]
\scriptsize
    \item
  \leavevmode
  \kern-\scriptspace
  \kern-\labelsep Estimates were obtained from the national health and nutrition examination survey (example 1, reliability of the error-prone blood pressure measurement: 0.95) \cite{NHANES} and the pregnancy day and assessment clinic study (example 2, reliability of the error-prone blood pressure measurement: 0.6) \cite{McCarthy2015EffectivenessHospital}
  \end{tablenotes}
\end{threeparttable}
\end{table}

\subsection{Review of regression calibration and simulation-extrapolation}\label{sec:reviewrcsimex}
In examples 1 and 2, the measurement error in the blood pressure measures was corrected by application of regression calibration and simulation-extrapolation. Regression calibration started by estimating the uncorrected effect of the error-prone first blood pressure measure on serum creatinine given age. Subsequently, the uncorrected estimate was multiplied by the estimated variance of the first error-prone blood pressure measurement given age, divided by the estimated variance of the first error-prone blood pressure measurement given age minus the measurement error variance. In examples 1 and 2, the measurement error variance was estimated using the replicate measurements, by estimating the within individual variance and averaging over all individuals. Alternatively, the measurement error variance could be informed by e.g. external data or expert knowledge. 

Simulation-extrapolation consists of two steps. In the simulation step, extra measurement error was added to the error-prone first blood pressure measurement of size 0.5, 1, 1.5, 2 times the measurement error variance. The measurement error variance was again estimated using the replicate measurements. Using these simulated blood pressure measurements with extra added measurement error, the effect of blood pressure on serum creatinine given age was estimated. This was repeated 100 times for each size of the measurement error variance and the newly obtained estimates were averaged. Then, in the extrapolation step, a model (e.g., linear, quadratic) was fitted trough the effect estimates for the varying sizes of the measurement error. The corrected effect estimate was then obtained by extrapolating the fitted model to the situation where the measurement error was equal to 0. For a visualisation of simulation-extrapolation, see e.g. \cite{Keogh2020STRATOSAdjustment}.

Ninety-five percent CI's for the corrected estimates were constructed by bootstrap resampling.

\section{SIMULATION STUDY}
We aimed to extend the simulation studies conducted by Perrier et al. \cite{Perrier2016Within-subjectStudies} and Batistatou et al. \cite{Batistatou2012PerformanceData}, by investigating the relative performance of regression calibration and simulation-extrapolation in scenarios different from those studied in the two former studies. The relative performance was studied in terms of bias, mean squared error, and confidence interval coverage of the true effect. A general description and motivation of the scenarios studied in our simulation study is provided in the following, the specific parameters set in our simulation study are explained in the subsequent subsection. 

Perrier et al. and Batistatou et al. assumed relatively large sample sizes (i.e., 3000 and 1000, respectively), only four different values for the reliability of the error-prone exposure (i.e., 0.2 and 0.6 in the study by Perrier et al. and 0.2, 0.5 and 0.8 in the study by Batistatou et al.) and a small coefficient of determination for the exposure-outcome model (i.e., 0.004 and 0.0625, respectively). In addition, Perrier et al. studied the effect of increasing the number of replicate measures on the performance of regression calibration and simulation-extrapolation. Yet, the replicate measures were pooled in the former study, thereby decreasing measurement error bias. Moreover, Perrier et al. and Batistatou et al. only examined models with a single independent variable. Our simulation study focused on relatively large reliability of the error-prone measurement (i.e., greater or equal to 0.625), motivated by the reliability of the blood pressure measures of our two motivating examples (i.e., 0.95 and 0.6 in example 1 and example 2, respectively). Additionally, the relative performance of regression calibration vs simulation-extrapolation was studied for small sample sizes (i.e., smaller or equal to 1000). In addition, the effect of a change in the coefficient of determination of the outcome model was tested. Furthermore, increasing the number of replicate measurements available was studied, without having the advantage of pooling the replicate measurements in our analysis. Lastly, multivariable models were studied. 

\subsection{Data generating mechanism}
Inspired by the data of our example of blood pressure and kidney function in pregnant women \cite{McCarthy2015EffectivenessHospital}, we assume the following data generating mechanisms for age, blood pressure (BP), error-prone blood pressure (BP$^*$) and creatinine: 
\begin{eqnarray*}
\textrm{Age} \sim \mathcal{N}(32,25), \quad \textrm{BP}|\textrm{Age}\sim \mathcal{N}(120 + \gamma\textrm{Age}, 50), \quad \textrm{BP}^*\sim \mathcal{N}(\textrm{BP}, \tau^2),\\
\textrm{and} \quad \textrm{Creatinine}|\textrm{BP, Age}\sim \mathcal{N}(30 + 0.2\textrm{BP} + 0.2\textrm{Age}, \sigma^2).
\end{eqnarray*}

The above defined data generating mechanism define that the error-prone blood pressure (BP$^*$) has random measurement error with measurement error variance equal to $\tau^2$. In our simulation study, a `base scenario’ was assumed and in the consecutive scenarios studied, we changed one of the four parameters in the data generating mechanisms (i.e., $\gamma, \tau^2$ or $\sigma^2$), the number of observations (i.e., $n$), or the number of replicate measures (i.e., $k$) (see Table \ref{tab:simsettings}). The parameters settings of the base scenario were inspired by our example of blood pressure and kidney function in pregnant women \cite{McCarthy2015EffectivenessHospital}. In the base scenario, sample size was 500, $\gamma = 0$, $\tau^2 = 30$ and $\sigma^2 = 100$ (Table \ref{tab:simsettings}). Further, we assumed that three replicate measures of the error-prone blood pressure measure were obtained in all individuals. From the parameter settings in the base scenario, it follows that the reliability of the error-prone measure is 0.625. The reliability is defined as the variance in the error-free blood pressure divided by the variance of the error-prone blood pressure. Further, in the base scenario, the R-squared of the outcome model is 0.03, and the attenuation due to measurement error of the effect of the error-prone blood pressure on creatinine (given age) is equal to the reliability, i.e., 0.625. 

In subsequent scenarios (see Table \ref{tab:simsettings}), we increased the reliability of the error–prone measure, by decreasing the variance of the error-prone measure (i.e., $\tau^2$). For completeness, 3 scenarios with low reliability were added. Subsequently, we changed the number of observations (see under ‘Sample size’ in Table \ref{tab:simsettings}). Next, we changed the number of replicate measures to study the effect of changing the precision of the estimation of the measurement error variance (see under `Number of replicates' in Table \ref{tab:simsettings}). Next, we decreased the residual error variance in the outcome model, $\sigma^2$ (see under ‘R-squared’ in Table \ref{tab:simsettings}). Decreasing the residual error in the outcome model, increases the R-squared of the outcome model. Lastly, we increased the reliability of the error-prone variable, by introducing a dependency between blood pressure and age, by changing $\gamma$ (see under ‘Covariate dependency’ in Table \ref{tab:simsettings}). By introducing an effect of age on blood pressure, the total variance of the error-free blood pressure increases. Consequently, the extra variability in the error-prone blood pressure measure due to measurement error is relatively smaller than in the base scenario. Hence, it seems as if the error-prone variable is more reliable, though the attenuation due to measurement error stays constant at a rate of 0.625. For each scenario, 5000 datasets were generated.

In each generated data set, the uncorrected effect was estimated using the first replicate measurement only. Subsequently, the corrected effect was estimated by application of regression calibration and simulation-extrapolation using the R package \texttt{mecor} \cite{Nab2021Mecor:Outcome} and \texttt{simex} \cite{Lederer1997AMCSIMEX}, respectively. The measurement error variance was estimated using the replicate measures. Ninety-five percent CI's of the corrected effects were constructed using bootstrap resampling. Performance of the three different analyses were evaluated in terms of bias, mean squared error (MSE), and the proportion of 95\% CIs that contained the true value of the estimand (coverage). Monte Carlo standard errors (MCSE) were calculated for all performance measures \cite{Morris2019UsingMethods}, using the R package \texttt{rsimsum} \cite{Gasparini2018Rsimsum:Studies}. All code used for the simulation study is publicly available via \texttt{www.github.com/LindaNab/simexvsmecor}.

\begin{table}[ht]
\centering
\caption{Simulation study settings}\label{tab:simsettings}
\begin{threeparttable}
\begin{tabular}{|l|l|l|l|l|l|}
\hline
     Scenarios & \multicolumn{5}{c|}{Parameters of Data Generating Mechanism\tnote{a}}\\
     \cline{2-6}
      & $\tau^2$ & $n$ & $k$ & $\sigma^2$ & $\gamma$ \\
     \hline
     Base & 30 & 500 & 3 & 100 & 0 \\
     \hline
     Reliability\tnote{b} & 200, & 500 & 3 & 100  & 0 \\
     & 100, & & & &  \\
     & 50, 25, & & & &  \\
     & 20, 15, & & & &  \\
     & 10, 5 & & & &  \\
     \hline
     Sample Size & 30 & 125, & 3 & 100 & 0 \\
      & & 250, & & & \\
      & & 1000, & & & \\
      & & 10,000 & & & \\
     \hline
     Number of & 30 & 500 & 2, 5, & 100 & 0 \\
     Replicates & & & 10 & & \\
      \hline
     R-squared\tnote{c} & 30 & 500 & 3 & 20, 5, 1 & 0 \\
     \hline
     Covariate & 30 & 500 & 3 & 100 & 1, 4, 8\\
     Dependency\tnote{c} & & & & &\\
     \hline
\end{tabular}
  \begin{tablenotes}[flushleft]
\scriptsize
    \item[a] $\tau^2$: measurement error variance of the error-prone blood pressure measurement; $n$: number of observations in the main study; $k$: number of replicate error-prone measurements; $\sigma^2$: residual variance of the outcome model; $\gamma$: association between blood pressure and age. The attenuation in the effect of blood pressure on creatinine due to measurement error is equal to $50/(50+\tau^2)$\\
    \item[b] Reliability is equal to $(25\gamma^2 + 50) / (25 \gamma^2 + 50 + \tau^2)$\\
    \item[c] R-squared is equal to $1 - \sigma^2 / (0.4\times 50 + 10 + \sigma^2)$\\
    \item[d] The effect of blood pressure on creatinine when age is not included in the model (crude model) is equal to $0.2 + 5\gamma / (25 \gamma^2 + 50)$
  \end{tablenotes}
\end{threeparttable}
\end{table}

\subsection{Results}
\begin{figure}[ht]
\centering
\begin{tabular}{c}
\includegraphics[]{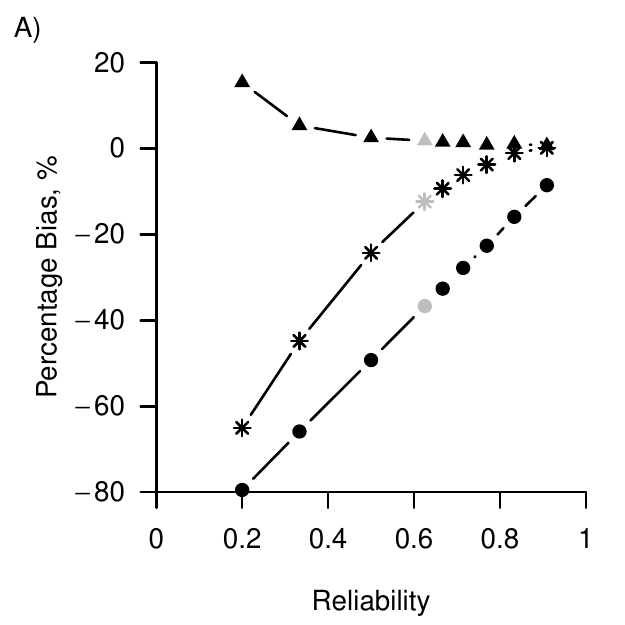}\\ \includegraphics[]{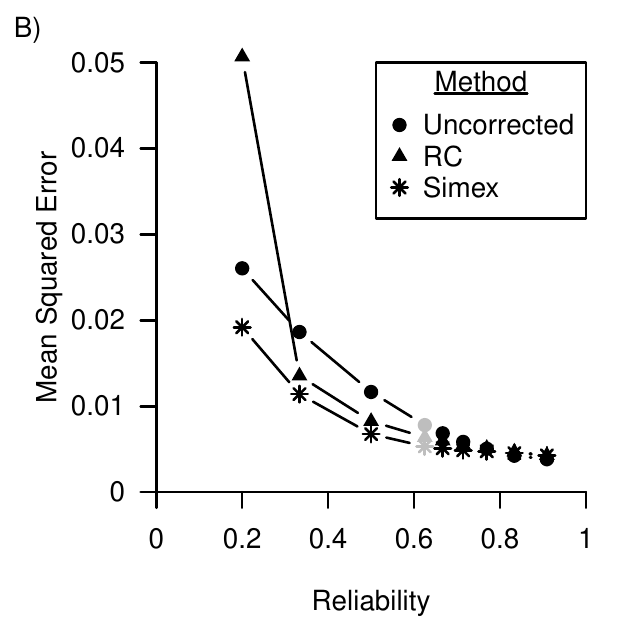}\\
\includegraphics[]{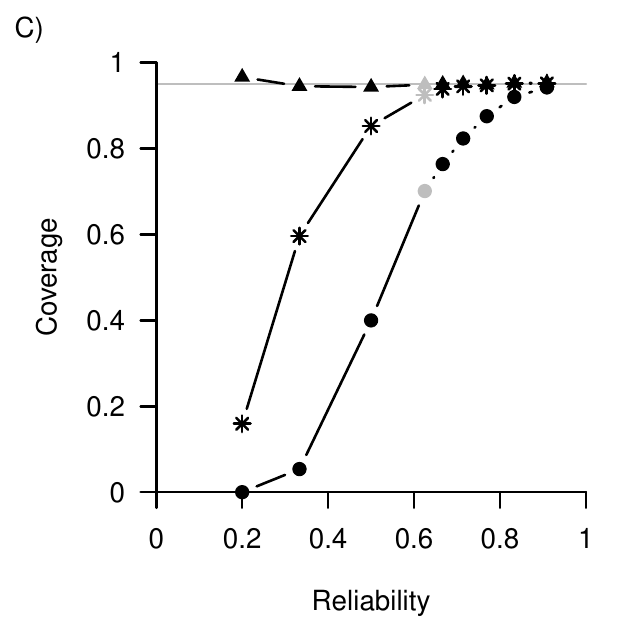}\\
\end{tabular}
\caption{Performance of regression calibration (RC), simulation-extrapolation (simex) and the analysis ignoring measurement error for varying values of the reliability of the error-prone measure in terms of A) percentage bias; B) mean squared error and C) coverage. For all three performance measures, Monte Carlo standard errors were smaller than 0.01 in all scenarios. The grey points indicate the base scenario where reliability is assumed 0.625.}\label{fig:reliability}
\end{figure}

\begin{figure}[ht]
\centering
\begin{tabular}{c}
\includegraphics[]{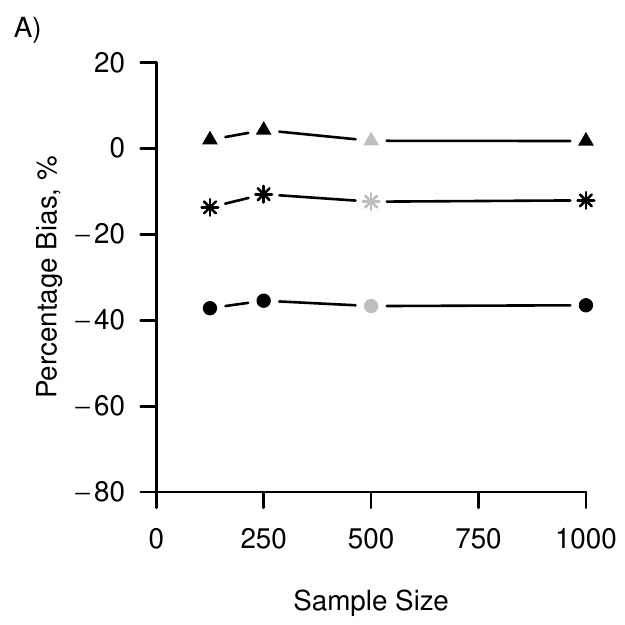}\\ \includegraphics[]{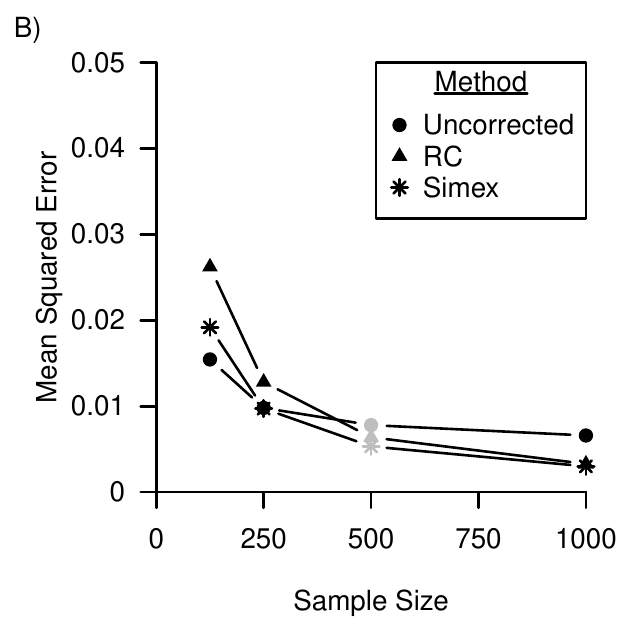}\\
\includegraphics[]{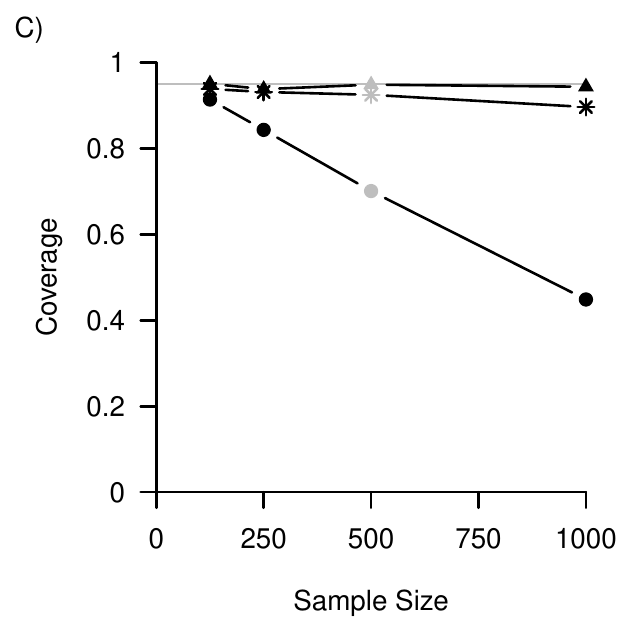}\\
\end{tabular}
\caption{Performance of regression calibration (RC), simulation-extrapolation (simex) and the analysis ignoring measurement error for varying sample sizes of the error-prone measure in terms of A) percentage bias; B) mean squared error and C) coverage. For all three performance measures, Monte Carlo standard errors were $<0.01$ in all scenarios. The grey points indicate the base scenario where sample size is assumed 500.}\label{fig:samplesize}
\end{figure}

\begin{figure}[ht]
\centering
\begin{tabular}{c}
\includegraphics[]{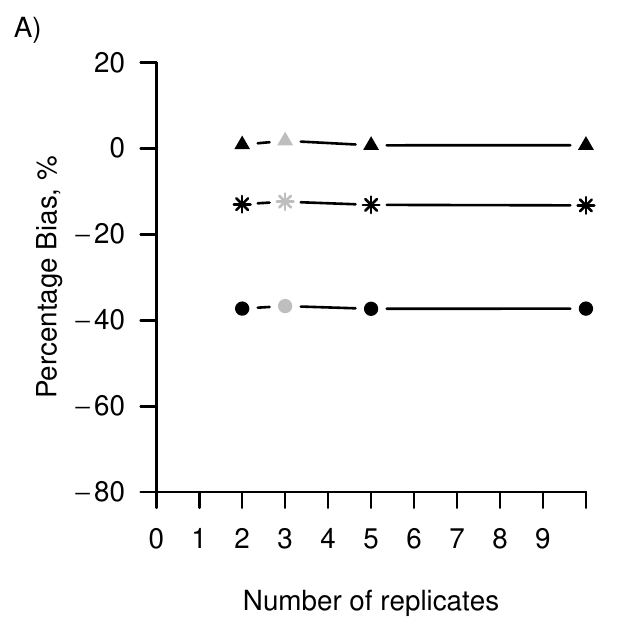}\\ \includegraphics[]{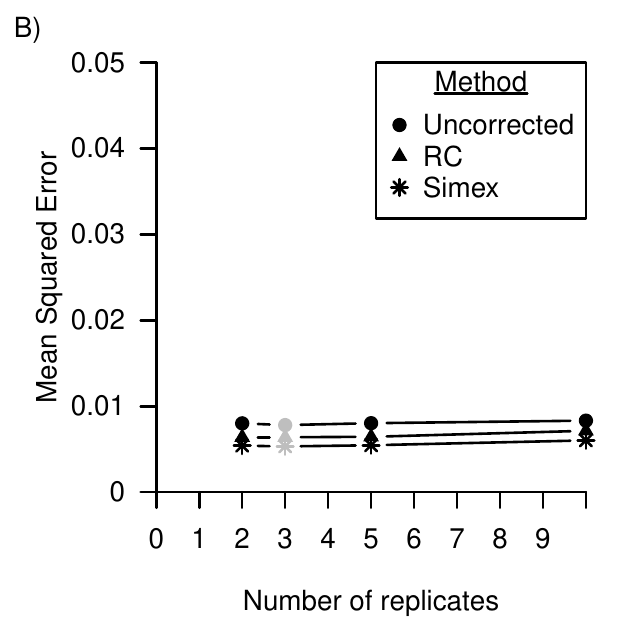}\\
\includegraphics[]{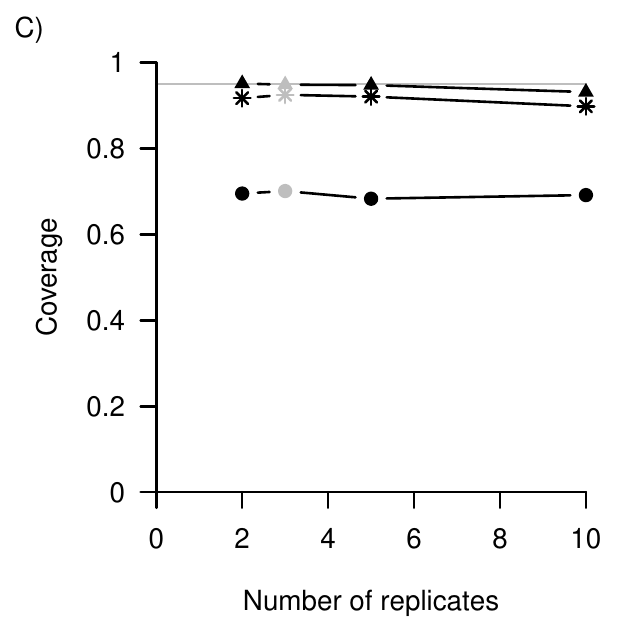}\\
\end{tabular}
\caption{Performance of regression calibration (RC), simulation-extrapolation (simex) and the analysis ignoring measurement error for varying number of replicates of the error-prone measure in terms of A) percentage bias; B) mean squared error and C) coverage. For all three performance measures, Monte Carlo standard errors were smaller than 0.01 in all scenarios. The grey points indicate the base scenario where the number of replicates is assumed 3.}\label{fig:nrep}
\end{figure}

\begin{figure}[ht]
\centering
\begin{tabular}{c}
\includegraphics[]{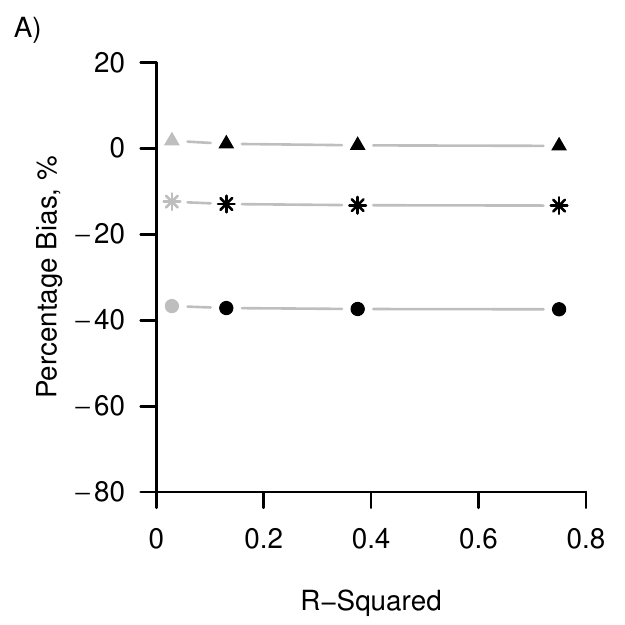}\\ \includegraphics[]{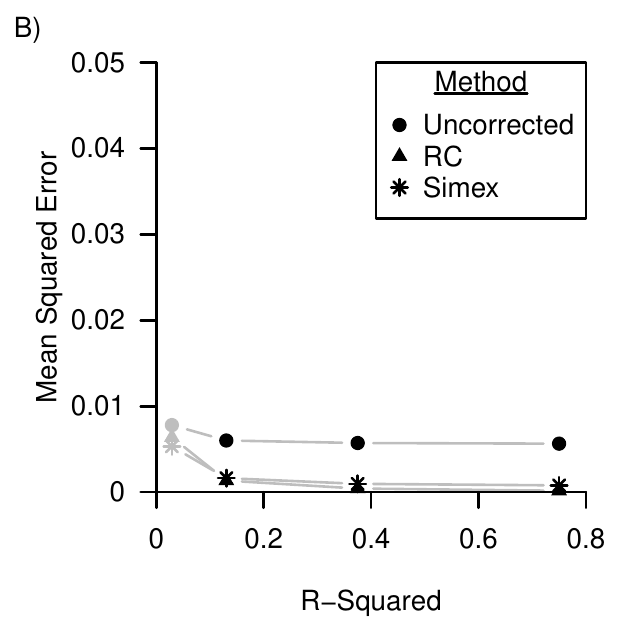}\\
\includegraphics[]{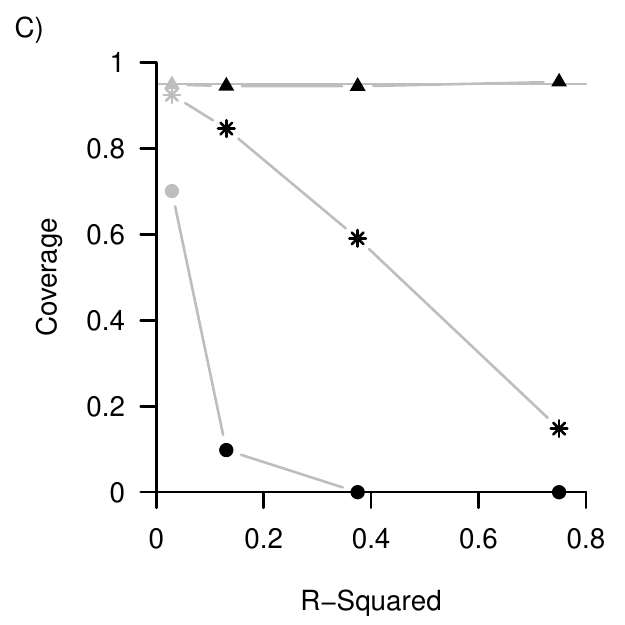}\\
\end{tabular}
\caption{Performance of regression calibration (RC), simulation-extrapolation (simex) and the analysis ignoring measurement error for varying R-Squared of the outcome model A) percentage bias; B) mean squared error and C) coverage. For all three performance measures, Monte Carlo standard errors were smaller than 0.01 in all scenarios. The grey points indicate the base scenario where R-Squared is assumed 0.03.}\label{fig:rsq}
\end{figure}

Figure \ref{fig:reliability} shows the percentage bias, MSE and confidence interval coverage for varying values of the reliability of the error-prone measure. The uncorrected analysis was biased for all values of the reliability, and the percentage bias decreased when reliability increased. Regression calibration provided unbiased results when reliability was greater or equal to $0.33$. Simulation-extrapolation provided biased results when reliability was smaller than $0.8$. MSE was lower for simulation-extrapolation than for the uncorrected and regression calibration corrected analysis when reliability was equal to 0.2, and similar to MSE of regression calibration otherwise. Coverage of the 95\% confidence intervals was at the nominal level for the regression calibration corrected analysis when reliability was greater than or equal to $0.33$, and for the simulation-extrapolation corrected analysis when reliability was greater than or equal to $0.625$.\\
Figure \ref{fig:samplesize} shows the percentage bias, MSE and confidence interval coverage for varying samples sizes of the main study. A sample size of 125, and 250 only increased percentage bias minimally compared to base scenario were sample size was 500. MSE was greater for smaller sample sizes, and MSE of the uncorrected analysis with a sample size of 125 was smaller than the regression calibration and simulation-extrapolation corrected analysis (0.015 vs 0.026 and 0.019, respectively, MCSE $<0.005$). Coverage was equal to the nominal level of 95\% for regression calibration for all sample sizes, and the uncorrected analysis was undercovered with coverage levels ranging between 91\% and 45\% (MCSE $<0.01$). Coverage of the 95\% confidence intervals of the simulation-extrapolation corrected analysis was close to the nominal level of 95\% except when sample size was 1000, in which case coverage was 90\% (MCSE $0.004$). A decline in confidence interval coverage for the simulation-extrapolation corrected analysis for larger sample sizes was confirmed by the scenario where sample size was 10,000, in which case coverage was 53\% (MCSE 0.007)(not shown in the plots in Figure \ref{fig:samplesize}).\\
Figure \ref{fig:nrep} shows that the number of replicates had no effect on the percentage bias, MSE and confidence interval coverage for varying number of replicates of the error-prone measure.\\
Figure \ref{fig:rsq} shows that R-squared had no effect on percentage bias, and only a minor decrease in MSE was found for increasing levels of R-Squared. In addition, Figure \ref{fig:rsq} shows that 95\% confidence interval coverage was around the nominal level for the regression calibration corrected analysis for all values of the R-squared. However, for the uncorrected and the simulation-extrapolation corrected analysis, confidence interval coverage decreased for increasing values of R-Squared. For R-Squared equal to 0.75, confidence interval coverage decreased to 15 \% and 0 \% (MCSE $<0.01$) for the simulation-extrapolation corrected and the uncorrected analysis, respectively.\\ 
In the scenarios where a dependency between the covariate age and the exposure error-free blood pressure was introduced by changing parameter $\gamma$ in the data generating mechanism, the reliability of the error-prone measure was respectively 0.71, 0.94 and 0.98. However, percentage bias, MSE and confidence interval coverage of the uncorrected and corrected analyses were equal to base scenario (the values in base scenario are shown in e.g. Figure \ref{fig:reliability}).

\clearpage

\section{SENSITIVITY ANALYSIS IN THE ABSENCE OF VALIDATION DATA}
In the two examples introduced in section 2, replicate measurements of the error-prone systolic blood pressure were available. Nevertheless, validation data in the form of replicate measurements may not always be available. When random measurement error in a covariate is suspected in the absence of such validation data, sensitivity analysis could be conducted using regression calibration or simulation-extrapolation. A general framework for conducting sensitivity analysis for measurement error is described here, where we assume that the input of the sensitivity analysis, i.e., the measurement error variance and its uncertainty, are obtained from literature or expert knowledge. Next, a distribution for the measurement error variance is assumed, e.g., a uniform, triangular, or trapezidiol distribution \cite{Lash2009ApplyingData}. Subsequently, regression calibration or simulation-extrapolation are applied to the data for measurement error correction, informed by the measurement error variance and its distribution. Finally the results of the application of measurement error correction are presented, and conclusions drawn about the sensitivity of the results to measurement error.

\subsection{Sensitivity analysis in the motivating example}
\begin{figure}[ht]
\centering
\begin{tabular}{ll}
\includegraphics[]{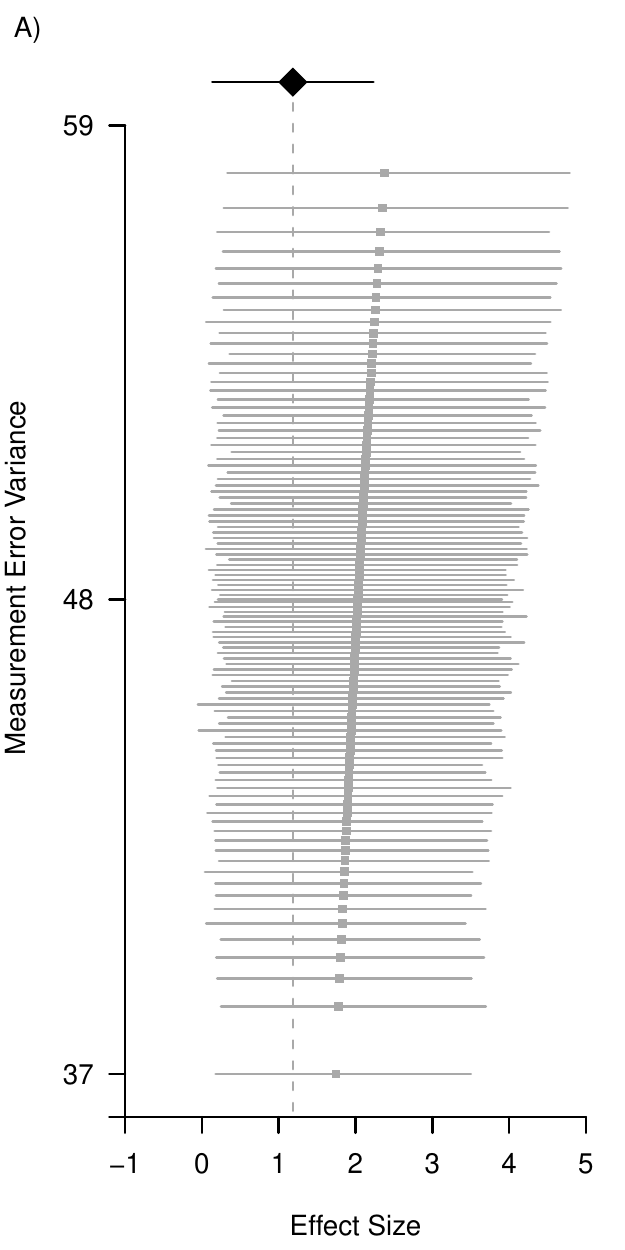} \includegraphics[]{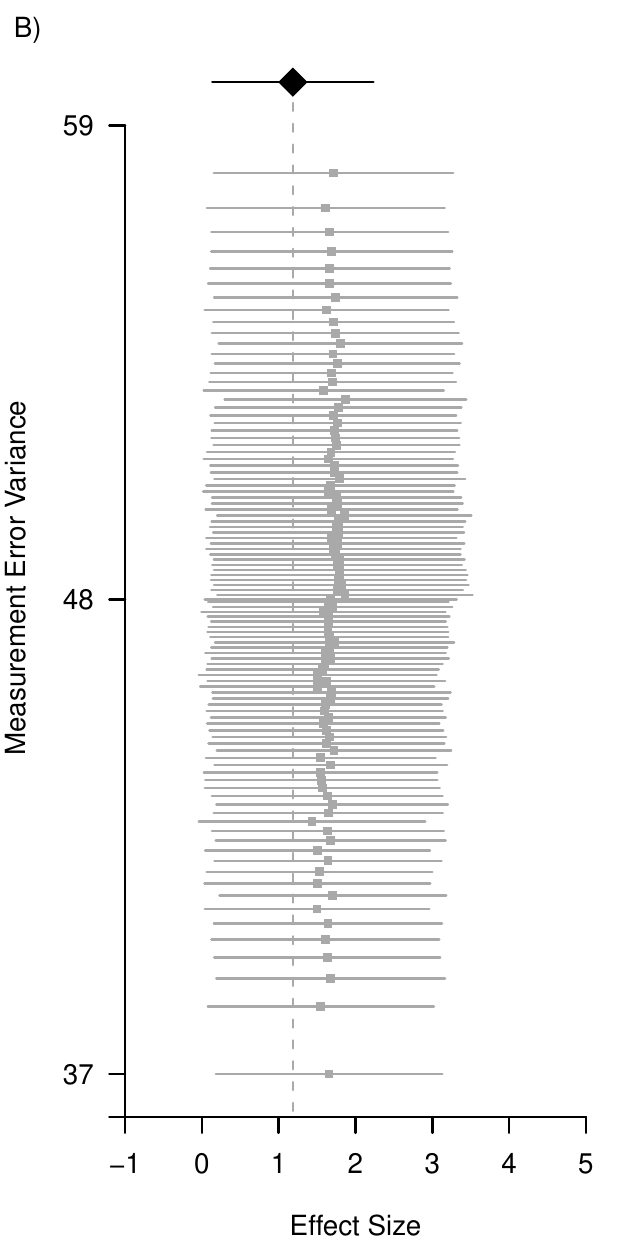}\\
\end{tabular}
\caption{Sensitivity analysis for the association between blood pressure and kidney function in pregnant women (example 2) by application of regression calibration (panel A) and simulation-extrapolation (panel B). The uncorrected association and 95\% confidence interval is depicted with a diamond and a solid black line, the measurement error corrected associations and 95\% confidence intervals are depicted with a square and a solid gray line. The distribution of the measurement error variance is triangular.}\label{fig:sensana}
\end{figure}

Suppose that in the example of the relation between blood pressure and kidney function in pregnant women discussed in section 2 (example 2), only the first systolic blood pressure measurement was available. A 10 mmHg increase in systolic blood pressure was associated with a 1.18 $\mu$mol/L (95\% CI 0.14-2.23) increase in serum creatinine. Measurement error, however, was suspected in the single systolic blood pressure measure and suppose the sensitivity of the results to the measurement error was studied. Suppose it was assumed that the variance of the measurement error in systolic blood pressure is equal to 48 mmHg, with a minimum of 37 mmHg and a maximum of 59 mmHg. Additionally, suppose a triangular distribution was assumed for the measurement error variance, meaning that most weight was put on 48 mmHg, and the weight was gradually reduced until it reached the assumed minimum and maximum level. The triangular distribution was sampled in accordance with Lash et al. \cite{Lash2009ApplyingData}.\\
Figure \ref{fig:sensana} shows the results of the application of regression calibration and simulation-extrapolation informed by the assumed triangular distribution. For regression calibration, a clear pattern was obtained in the sensitivity analysis plot shown in Figure \ref{fig:sensana}. The corrected effect estimates increased for larger values of the measurement error variance, with the effect estimates ranging from 1.75 - 2.38, with a median of 2.03. In addition, the associated lower limits of the confidence intervals consistently suggest an association between blood pressure and creatinine. In comparison, simulation-extrapolation did not show a clear pattern in the corrected effect estimates. The corrected effect estimates ranged from 1.43 - 1.88, with a median of 1.70. Figure \ref{fig:sensana} shows that the sampling variability that is inherent to simulation-extrapolation causes more variability in the effect estimates compared to the variability due to measurement error. Nevertheless, the lower limits of the associated confidence intervals again consistently suggest an association between blood pressure and creatinine.

\section{DISCUSSION}
This paper compared regression calibration and simulation-extrapolation for conducting a sensitivity analysis for random measurement error in an exposure variable. A simulation study showed that with correct assumptions about the measurement error variance, the regression calibration corrected analysis was generally unbiased. However, the uncorrected and simulation-extrapolation corrected analysis were generally biased, with higher bias for lower reliability of the error-prone exposure. The uncorrected analysis showed worse efficiency in terms of mean squared error over the two corrected analyses, expect for small sample sizes (i.e., 125). Compared to regression calibration, the simulation-extrapolation corrected analysis showed similar efficiency in terms of mean squared error, except for low reliability (i.e., 0.2 and a sample size of 500) and small sample size (i.e., 250 and 125, and a reliability of 0.625). In addition, regression calibration showed confidence interval coverage of levels close to 95\% in all scenarios, while the confidence intervals of the simulation-extrapolation corrected analysis and uncorrected analysis were often undercovered. What is more, an increment of the R-squared value of the outcome model showed a large decrease in confidence interval coverage of the true effect for the uncorrected and simulation-extrapolation corrected analysis.

The results of our simulation study were in line with the results of two previous simulation studies: the corrected analyses showed lower percentages bias compared to the uncorrected analysis and the simulation-extrapolation corrected analysis showed higher percentage bias compared to regression calibration \cite{Perrier2016Within-subjectStudies, Batistatou2012PerformanceData}. However, important differences were observed. First, simulation-extrapolation showed a small gain in efficiency over regression calibration in some settings, which was not found in the previous simulation studies. The sample sizes for which this gain in efficiency for simulation-extrapolation was observed (i.e., 125, 250 and 500) were smaller than those assumed by Perrier et al. \cite{Perrier2016Within-subjectStudies} and Batistatou et al. \cite{Batistatou2012PerformanceData} (i.e., 3000 and 1000, respectively), which may explain the found difference. Second, our simulation showed no effect of the number of replicates on bias. While the simulation study by Perrier et al. showed that an increasing number of replicates reduced bias in the corrected analyses \cite{Perrier2016Within-subjectStudies}. This difference is explained by the fact that in the study by Perrier et al., the replicate measures were pooled before applying measurement error correction. By pooling the replicate measures with random measurement error, the measurement error variance is reduced. Therefore, bias decreased in the study by Perrier et al. with the availability of more replicate measures. This effect however is solely due to pooling the replicates measures and not due to a more precise estimate of the measurement error variance, as was shown by our results. Of note, when replicate measurements are obtained, pooling the replicate measurements is always preferred above only using the replicate measurements to estimate the measurement error variance.

Our simulation study showed that percentage bias in the uncorrected analysis was equal to 1 minus the reliability of the error-prone measure times 100, in line with theory \cite{Frost2000CorrectingVariable., Hutcheon2010RandomBias}. The reliability of an error-prone measure equals the variance of the error-free measure divided by the variance of the error-prone measure. For example, in Figure \ref{fig:reliability}, bias in the uncorrected analysis was equal to 80\% for a reliability equal to 0.2. The uncorrected effect estimate is equal to 0.2 times the estimand 0.2, i.e., 0.04. From that, it follows that the bias is equal to $0.2-0.04 = 0.16$, which is 80\% of 0.2. It is, however, important to note that the percentage bias is not equal to 1 minus the reliability of the error-prone measure when the total variance of the error-free measure depends on a covariate that is also included in the outcome model. For example, in our simulation study, the association between creatinine and systolic blood pressure given age was estimated. When a dependency between systolic blood pressure and age was introduced, the reliability increased to a maximum of 0.98 while the percentage bias in the uncorrected analysis was constant at 62.5\%. A formula for the attenuation in the effect estimate due to random measurement error in multivariable models can be found in e.g. \cite{Carroll2006}.

In our simulation study, the measurement error variance used to correct for the measurement error was estimated using replicate measures. However, we assumed that these replicate measures were solely available to estimate the measurement error variance, to mimic a setting where such validation data is not available but with unbiased estimation of the measurement error variance. In future studies, this work could be extended to settings where the measurement error is estimated with bias, and to settings where the measurement error model is misspecified. Also systematic measurement error, which was not the topic of this study, could be considered in future work. 

In the example presented in section 4, the five steps of a sensitivity analysis for exposure measurement error were described: 1) quantify the measurement error variance and its uncertainty; 2) specify the distribution of the measurement error variance; 3) perform measurement error correction by means of regression calibration or simulation-extrapolation; 4) visualise the results, and 5) draw conclusions. A sensitivity analysis using regression calibration showed that the higher the measurement error variance, the more the corrected effect estimate departs from the null, which is in line with the literature \cite{Frost2000CorrectingVariable., Hutcheon2010RandomBias}. In the sensitivity analysis using simulation-extrapolation, the variability in the corrected effect estimates due to the sampling variability inherent to simulation-extrapolation exceeded the variability in the corrected effect estimates due to the assumed measurement error variance. Since regression calibration was generally shown unbiased and simulation-extrapolation was generally shown biased, the application of regression calibration for conducting sensitivity analyses is recommended. Moreover, the advantage in efficiency of simulation-extrapolation over regression calibration in some settings does play a less important role in sensitivity analysis than its disadvantage in performance due to bias. 

In conclusion, in the absence of validation data, regression calibration and simulation-extrapolation are suitable for conducting sensitivity analyses for random measurement error. Particularly, regression calibration is generally preferred over simulation-extrapolation due to its unbiasedness in most settings. To guide sensitivity analysis for random measurement error, a sensitivity analysis framework was provided.

\clearpage

\bibliographystyle{unsrt}
\bibliography{ref.bib}

\end{document}